\documentclass{elsart1p}

\usepackage{amssymb,graphicx}

\begin{document}

\begin{frontmatter}



\title{Multiloop Bubbles for hot QCD}

%\author{E. Bejdakic \thanksref{label2}}
% \address[label2]{ervin@physik.uni-bielefeld.de}

\author{E. Bejdakic\thanksref{label1}}
%\author{ Y.Schr\"{o}der \ead{yorks@physik.uni-bielefeld.de} \thanksref{label1}}
%\corauthref{cor1}\thanksref{label2}
\address[label1]{Fakult\"{a}t f\"{u}r Physik, Universit\"{a}t Bielefeld, 33501 Bielefeld, Germany}
%\address[label2]{yorks@physik.uni-bielefeld.de}
%\corauth[cor1]{hhhhhhhhhh}
\begin{abstract}
% Text of abstract
In this article, we present analytical expansion results of two single mass scale four-loop vacuum integrals in $d=3-2\epsilon$ dimensions. After finding hypergeometric representations with half-integer coefficients, we use algorithms which we implemented in FORM to expand these in terms of nested sums. 
\end{abstract}

\end{frontmatter}

% main text
\section{Introduction}
%The euclidian Lagrangian of QCD is given by 
%\begin{equation}
%        \mathcal{L}_{QCD}=\frac{1}{4g^2}F_{\mu\nu}^{a}F_{\mu\nu}^a+\bar{\psi}\gamma_{\mu}D_{\mu}\psi
%\end{equation}
%One interesting observable is the pressure, give by: 
%\begin{equation}
%p_{QCD}(T)=\lim\limits_{V \to\infty}\ln\int\mathcal{D}[A_{\mu}^a,\psi,\bar{\psi}]
%         \exp[-\int\limits_{0}^{1/T}d\tau \int d^{3-2\epsilon}x\mathcal{L}_{QCD}]
%\end{equation}
%The pressure is of importance, for example, in cosmology, where it determines the %cooling rate of the universe [ref] according to
% \begin{equation}
%\partial_tT=-\frac{\sqrt{24\pi}}{m_{pl}}\cdot\frac{\sqrt{e(T)}}{\partial_T\ln s(T)}
%\end{equation}
%where $ e = Ts-p $ is the energy density and $ s = \partial_Tp $ the entropy density of the universe.
%The temperature enters in the eq.(2) as upper bound in the integral over $\tau$ making it compact, thus when going to Fourier space in order to compute pressure in weak coupling expansion we will have three dimensional integrals and one sum instead of four dimensional integrals like in zero temperature quantum field theory. The structure of the weak coupling expansion series is not analytical in $g^2$.
%The reason for this is that hot QCD is a multiscale system. One way of dealing with this problem is using effective field theory approach. In this approach the different scales are separated within different effective theories. You can imagine it this way;
The series of the pressure of QCD in weak coupling expansion is not analytical in the coupling $g^2$. The reason for this is that hot QCD is a multiscale system. One way of dealing with this problem is using effective field theory approach \cite{Appelquist:1981vg,Ginsparg:1980ef,Braaten:1995cm}. In this approach the different scales are separated within different effective theories.
At temperatures well above $T_c$ all the relevant information of QCD is contained within a three dimensional effective field theory, the so-called EQCD. For the specific problem mentioned above, EQCD
%\section{EQCD Masters} 
% we now apply the whole perturbative formalism of QFT, such as diagram generation (with classification, scalarization) and reduction of the huge number of diagrams generated to so called master integrals using  Integration  By  Parts method \cite{Chetyrkin:1981qh}. These procedures
has the following set of master integrals up to four loops \cite{Schroder:2002re}, obtained using methods from \cite{Chetyrkin:1981qh,Tkachov:1981wb,Laporta:2001dd}: \\
\begin{equation}
 \parbox{1.3cm}{\includegraphics[width=0.8\textwidth]{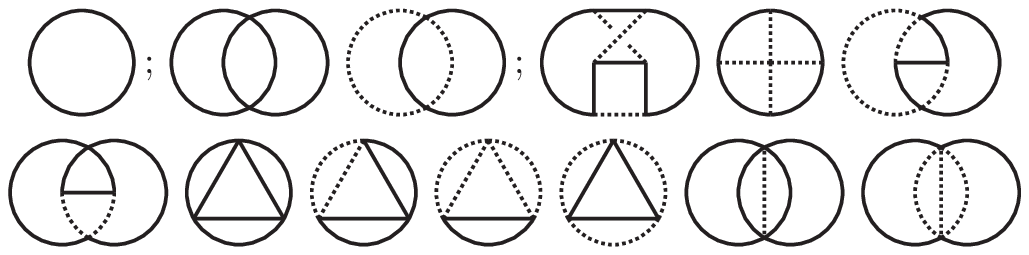}}
\end{equation} 
%\begin{picture}(800,230)
%  \put(0,0){\centerline{\includegraphics[width=0.3\textwidth]{master_QED}}}
%\end{picture}
%\begin{center}
%\begin{minipage}{30cm}
%  {\large{\centerline{\it Master integrals of EQCD up to four loops\\}}}
%\end{minipage}
%\end{center}
The full/dotted lines stand for a massive/massless scalar propagators. These {\em Masters} need to be computed. There are several methods for doing this; among the most frequently used are solving the corresponding difference equations \cite{Laporta:2001dd}, as well as the Mellin-Barnes method \cite{Boos:1990rg,Smirnov:2004ym}. Here we will only be using Mellin-Barnes method for obtaining hypergeometric representation of some Masters, while as a second step we use nested sums \cite{Vermaseren:1998uu, Moch:2001zr} for expanding these hypergeometric functions analytically around $d=3-2\epsilon$ dimensions.

\section{Mellin-Barnes method}
The Mellin-Barnes method for evaluating massive Feynman integrals consists of writing down a massive propagator in terms of a massless one \cite{Boos:1990rg}:
\begin{equation}
        \frac{1}{(k^2+m^2)^{\beta}}=\frac{1}{(k^2)^{\beta}\Gamma(\beta)2\pi i}
	\int\limits_{-i\infty}^{i\infty}ds \left(\frac{m^2}{k^2}\right)^s
	\Gamma(-s)\Gamma(\beta+s) \; ,
\end{equation}
where the contour of integration is chosen such that the poles of $\Gamma(-s)$ are to the right of the contour.
One simple example is the one-loop integral of eq.(3), which using the Mellin-Barnes method gives us a massless one-loop integral which in turn can be evaluated in terms of Gamma functions. At the end one can perform the complex Mellin-Barnes integral closing the contour to the right, obtaining:
%\begin{equation}
%	 I(a,b;m)\equiv \int\frac{d^{{d}}p}{(p^2)^a((k-p)^2+m^2)^b} \stackrel{a,b=1}{\equiv}  \parbox{0.3cm}{\includegraphics[width=0.1\textwidth]{selfEdots}}
%\end{equation}

\begin{eqnarray}
 \int\frac{d^{{d}}p}{(p^2)^a((k-p)^2+m^2)^b} &=&  \pi^{\frac{{d}}{2}}(m^2)^{\frac{{d}}{2}-a-b}\frac{\Gamma(\frac{{d}}{2}-a)}
{\Gamma(a)\Gamma(b)}\nonumber\\
& &\sum\limits_{j=0}^{\infty}\frac{1}{j!}\left(\frac{k^2}{m^2}\right)^j \frac{\Gamma(a+j)\Gamma(a+b-\frac{{ d}}{2}+j)}{\Gamma(\frac{{d}}{2}+j)} \; .
\end{eqnarray}

%Inserting the Mellin-Barnes representation for the massive propagator gives us the corresponding massless integral and an additional Mellin-Barnes integration. The massless integral can be done analytically in terms of Gamma functions:
%\begin{equation}
%\int\frac{d^{d}p}{(p^2)^a((k-p)^2)^b}=\pi^{\frac{{d}}{2}}(k^2)^{\frac{{d}}{2}-a-b}
%\frac{\Gamma(\frac{{d}}{2}-a)\Gamma(\frac{{d}}{2}-b)\Gamma(a+b-\frac{{d}}{2})}
%{\Gamma(a)\Gamma(b)\Gamma({d}-a-b)}
%\end{equation}
%Using this result all what is left to do is the Mellin-Barnes integral, which by closing to contour to the right and using properties of the Gamma functions gives us:
%\begin{eqnarray}
%I(a,b;m)&=&\pi^{\frac{{d}}{2}}(m^2)^{\frac{{d}}{2}-a-b}\frac{\Gamma(\frac{{d}}{2}-a)}
%{\Gamma(a)\Gamma(b)}
%\sum\limits_{j=0}^{\infty}\frac{1}{j!}\left(\frac{k^2}{m^2}\right)^j \frac{\Gamma(a+j)\Gamma(a+b-\frac{{ d}}{2}+j)}{\Gamma(\frac{{d}}{2}+j)}\\
%&=&\pi^{\frac{{d}}{2}}(m^2)^{\frac{{d}}{2}-a-b}\frac{\Gamma(\frac{{d}}{2}-a)\Gamma(a+b-\frac{{d}}{2})}
%{\Gamma(\frac{{d}}{2})\Gamma(b)} \hspace{0cm}_2F_1(a,a+b-\frac{{d}}{2};\frac{{d}}{2};\frac{k^2}{m^2})
%\end{eqnarray}

What is left to do now is the sum over Gamma functions. Fortunately, there are methods developed \cite{Moch:2001zr, Weinzierl:2004bn} for expanding in $\epsilon$ such hypergeometric sums in terms of objects called nested sums. Note that in the specific example above, for $d=4-2\epsilon$ one would have integer valued coefficients in $\Gamma$-functions, while for $d=3-2\epsilon$ one is left with half-integer valued coefficients.

\section{Nested Sums}

S-Sums \cite{Vermaseren:1998uu, Moch:2001zr} are defined recursively as: $S(n)=1$ for $n>0$ , $S(n)=0$ for $n\le0$ and
%\[S(n)      =  \left\{ \begin{array}{c}
%1 : n>0 \\ 0 : n\le0 \end{array}\right. \]
\begin{eqnarray}
S(n;m_1,\ldots,m_k;x_1,\ldots,x_k)  =  \sum_{i=1}^n \frac{x_1^i}{i^{m_i}}S(i;m_2,\ldots,m_k;x_2,\ldots,x_k)\; .
\end{eqnarray}
%It can be shown \cite{Moch:2001zr} that these nested sums form an algebra, the so called Hopf algebra. 
Gamma functions can be naturally expanded in terms of S-sums, according to:
\begin{equation}
\frac{\Gamma(j+1+\epsilon)}{\Gamma(1+\epsilon)} =
\Gamma(j+1)\exp\left(-\sum\limits^{\infty}_{k=1}\epsilon^k \frac{(-1)^k}{k}S(j;k;1)\right)\; .
\end{equation}
So if we can represent Feynman integrals in terms of one infinite sum over Gamma functions, we can expand the Gamma functions in terms of S-Sums, use their algebra \cite{Vermaseren:1998uu,Moch:2001zr} to reduce products of S-sums to a linear combination of S-sums and finally do the last infinite sum over S-sums. This last sum gives S-sums with argument infinity, which are, per definition, multiple polylogarithms (MPL)\cite{Goncharov:1998}. In eq.(6,7) MPL simplify further to multiple $\zeta$-values \cite{Borwein:1999js}. There are packages for the expansion of hypergeometric functions with integer valued coefficients \cite{Moch:2005uc,Huber:2005yg,Weinzierl:2002hv} and of some classes of functions with half-integer coefficients \cite{Huber:2007dx}. However, for some of our Masters, in particular the one of eq.(6), these are not sufficient, what prompted us to implement the relevant algorithms of \cite{Weinzierl:2004bn}.

\section{Applications}
Here we apply the methods discussed in previous sections to two four-loop single mass scale Feynman integrals from EQCD. We first apply the Mellin-Barnes method and get hypergeometric representations. These consists in the case of eq.(7) of one hypergeometric function, which in $d=3-2\epsilon$ is given by $_4F_3\Big(1,1-\epsilon, 2\epsilon, \frac{1}{2}+\epsilon;2-2\epsilon,1+\epsilon,\frac{3}{2}-\epsilon;1\Big)$ and in case of eq.(6) of nine different hypergeometric functions, of which the highest one is $_5F_4\Big(1,1-\epsilon, \frac{3}{2}+\epsilon, \frac{3}{2}+\epsilon,2+2\epsilon;2-2\epsilon,\frac{3}{2}-\epsilon,\frac{5}{2}+\epsilon,2+\epsilon;1\Big)$.
As mentioned above, we implemented the algorithms for expansion in $\epsilon$ of half-integer valued balanced hypergeometric functions in FORM \cite{Vermaseren:2000nd}. Hypergeometric representations in $d$ dimensions and expansions in $d=4-2\epsilon$ will be published elsewhere \cite{Bejdakic:prep}. Note that both integrals are normalized by the appropriate power of the massive one-loop tadpole integral. \\
%\begin{eqnarray}
%\frac{\parbox{1cm}{\includegraphics[width=0.09\textwidth]{651}}}{J^4}\hspace{0.7cm}&=&\frac{1}{24}\Big(\pi ^2-12 \ln^{2}2\Big)+\epsilon\Big(5 \ln^{2}2+\ln^{3}2-\frac{\pi ^2}{12}(5-\ln512)\nonumber\\
%&-&4\, \zeta_3\Big) +\epsilon^2\Big(30\, a_4-\frac{\pi ^4}{144}-26 \ln^{2}2-10 \ln^{3}2+\frac{\ln^{4}2}{12}\nonumber\\
%&-&\frac{1}{12} \pi ^2 (-26+90 \\ln2+23 \ln^{2}2)+40\, \zeta_3-\frac{21}{4}\,\zeta_3 \\ln2 \Big) +\mathcal{O}(\epsilon^3)
%\end{eqnarray}

\begin{eqnarray}
\frac{\parbox{1cm}{\includegraphics[width=0.09\textwidth]{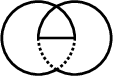}}}{\hspace{0.2cm}\Big(\parbox{0.6cm}{\includegraphics[width=0.05\textwidth]{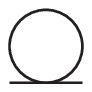}}\hspace{0.6mm}\Big)^4}\hspace{0.7cm}&=&\vspace{-8mm}\frac{1}{4} \, \zeta_2-\frac{1}{2} \ln^{2}2+ \epsilon(-4\, \zeta_3 - \frac{5}{2}\, \zeta_2 + \frac{9}{2}\ln2\, \zeta_2 + 5 \ln^{2}2 + \ln^{3}2 )\nonumber\\[-4mm]
&+& \epsilon^2( 30\,a_4 + 40\, \zeta_3 + 13\, \zeta_2 - \frac{1}{4}\, \zeta_2^2 - \frac{21}{4}\ln2 \, \zeta_3 - 45\ln2 \, \zeta_2 - 26 \ln^{2}2\nonumber\\
&-& \frac{23}{2} \ln^{2}2 \, \zeta_2 - 10 \ln^{3}2 +  \frac{1}{12} \ln^{4}2 )+ \epsilon^3 (-28\, a_5- \frac{2103}{16}\, \zeta_5 - 300\, a_4 \nonumber\\
&-& 208 \, \zeta_3- 54 \, \zeta_2 - 13 \, \zeta_2 \, \zeta_3 + \frac{5}{2} \, \zeta_2^2+ 28 \ln2\, a_4+ \frac{105}{2} \ln2 \, \zeta_3 + 234 \ln2 \, \zeta_2 \nonumber\\
&+& \frac{361}{5} \ln2 \, \zeta_2^2 + 108 \ln^{2}2 + \frac{213}{4}\ln^{2}2 \, \zeta_3 + 115 \ln^{2}2 \, \zeta_2 + 52 \ln^{3}2 \nonumber\\
&-& \frac{14}{3} \ln^{3}2 \, \zeta_2 - \frac{5}{6} \ln^{4}2 + \frac{12}{5} \ln^{5}2 ) + \mathcal{O}(\epsilon^4)
%\epsilon^4(-\frac{9}{8}- 278 \,s6 + 24 a_6+ 280 a_5 \nonumber\\
%&+& \frac{12531}{8} \, \zeta_5 + 1560 a_4 + 1632 \, \zeta_3 + \frac{4325}{16}\, \zeta_3^2 - 552 \, \zeta_2 - 12 \, \zeta_2 a_4 + 604 \, \zeta_2 \, \zeta_3 \nonumber\\
%&-& 1063 \, \zeta_2^2 - \frac{34901}{210} \, \zeta_2^3 - \frac{3319}{8} \ln2 - 184 \ln2 a_5 + \frac{1755}{8}\ln2 \, \zeta_5 \nonumber\\
%&-& 280 \ln2 a_4 - 393 \ln2 \, \zeta_3 + 316 \ln2 \, \zeta_2 - \frac{317}{2} \ln2 \, \zeta_2 \, \zeta_3
%          + \frac{2324}{5} \ln2 \, \zeta_2^2 \nonumber\\
%&-& 432 \ln^{2}2 - 92 \ln^{2}2 a_4 - \frac{1065}{2} \ln^{2}2 \, \zeta_3 - 1150 \ln^{2}2 \, \zeta_2 - 175 \ln^{2}2 \, \zeta_2^2 \nonumber\\
%&-& 216 \ln^{3}2 - \frac{199}{2} \ln^{3}2 \, \zeta_3 + \frac{172}{3} \ln^{3}2 \, \zeta_2 - 33 \ln^{3}2 \, \zeta_2 \, \zeta_3 + 16 \ln^{3}2 \, \zeta_2^2 \nonumber\\
%&+& \frac{13}{3} \ln^{4}2 + \frac{73}{3} \ln^{4}2 \, \zeta_2 - 24 \ln^{4}2 \, \zeta_2^2 - 24 \ln^{5}2 + 8 \ln^{5}2 \, \zeta_2 - \frac{133}{45} \ln^{6}2 \nonumber\\
%&-& \frac{16}{3} \ln^{6}2 \, \zeta_2 )+\mathcal{O}(\epsilon^5)
\end{eqnarray}

\begin{eqnarray}
\frac{\parbox{1cm}{\includegraphics[width=0.09\textwidth]{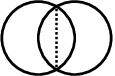}}} {\hspace{0.2cm}\Big(\parbox{0.6cm}{\includegraphics[width=0.05\textwidth]{tadpole}}\hspace{0.6mm}\Big)^4}\hspace{0.5cm}&=&
\frac{7}{4\epsilon}+7-8\ln2+\epsilon(49+16\,\zeta_2-32\ln2+16\ln^{2}2)\nonumber\\[-4mm]
&+&\epsilon^2(308-108\,\zeta_3+64\,\zeta_2-224\ln2-64\,\zeta_2\ln2+64\ln^{2}2-\frac{64}{3}\ln^{3}2)\nonumber\\
&+&\epsilon^3(1904+128\, a_4-432\,\zeta_3+448\,\zeta_2+\frac{412}{5}\,\zeta_2^2-1408\ln2+544\,\zeta_3\ln2\nonumber\\
&-&256\,\zeta_2\ln2+448\ln^{2}2+96\,\zeta_2\ln^{2}2-\frac{256}{3}\ln^{3}2+\frac{80}{3}\ln^{4}2+426\,\zeta_4)\nonumber\\
&+&\epsilon^4(11648+512\,a_5-3212\,\zeta_5 + 512\,a_4 - 3024\,\zeta_3 + 2816\,\zeta_2 - 1088\,\zeta_2\,\zeta_3\nonumber\\
&+& \frac{1648}{5}\,\zeta_2^2 - 8704\ln2 + 2176\ln2\,\zeta_3 - 1792\ln2\,\zeta_2
          - \frac{1648}{5}\ln2\,\zeta_2^2\nonumber\\
&+& 2816\ln^{2}2 - 1088\ln^{2}2\,\zeta_3 + 384\ln^{2}2\,\zeta_2 - 1792/3\ln^{3}2 - 128\ln^{3}2\,\zeta_2\nonumber\\
&+& \frac{320}{3}\ln^{4}2 - 64/3\ln^{5}2 + 1704\,\zeta_4 - 1704\,\zeta_4
         \ln2 )\nonumber\\
&+& \epsilon^5(70784-256 \,s_6+2048\, a_6+2048\, a_5-12848\, \zeta_5 + 3584 \,a_4\nonumber\\
&-& 19008 \,\zeta_3+ 3768\,\zeta_3^2 + 17408\,\zeta_2 - 4352\,\zeta_2\,\zeta_3 + \frac{11536}{5}
         \,\zeta_2^2 + \frac{7968}{35}\,\zeta_2^3\nonumber\\
&-& 53248\ln2+ 13344\ln2\,\zeta_5 + 15232\ln2\,\zeta_3 - 11264\ln2\,\zeta_2\nonumber\\
&+& 4352\ln2\,\zeta_2\,\zeta_3- \frac{6592}{5}\ln2\,\zeta_2^2+ 17408\ln^{2}2 - 4352\ln^{2}2
         \,\zeta_3\nonumber\\
&+& 2688\ln^{2}2\,\zeta_2+ \frac{3296}{5}\ln^{2}2\,\zeta_2^2 - \frac{11264}{3}\ln^{3}2 + \frac{4352}{3}\ln^{3}2\,\zeta_3- 512\ln^{3}2\,\zeta_2\nonumber\\
&+& \frac{2240}{3}\ln^{4}2 + 128\ln^{4}2\,\zeta_2 - \frac{256}{3}\ln^{5}2 + \frac{128}{9}
         \ln^{6}2+11928\,\zeta_4\nonumber\\
&+& 5592\,\zeta_4\,\zeta_2 - 6816\,\zeta_4\ln2 + 3408\,\zeta_4\ln^{2}2 + 11146\,\zeta_6 )+\mathcal{O}(\epsilon^6)\; ,
\end{eqnarray}
where $s_6 = \sum\limits_{i_1=1}^{\infty}\frac{(-1)^{i_1}}{i_1^5}\sum\limits_{i_2=1}^{i_1}\frac{(-1)^{i_2}}{i_2} = 0.987441...$  and $a_i=\mbox{Li}_i\Big(\frac{1}{2}\Big)$.

\section{Summary}
We implemented in FORM algorithms for expanding hypergeometric functions, including expansion around half-integer balanced values. As an application we showed expansions in $d=3-2\epsilon$ of two four-loop bubbles needed in the context of hot QCD.
%\section{Acknowledgment}
%This work has been supported by DFG under grant GRK 881.
% \appendix

% \section{}
% \label{}

\end{document}